\documentclass[pra, twocolumn, showpacs, preprintnumbers, amsmath, amssymb, superscriptaddress, aps]{revtex4}
\usepackage{amssymb}
\usepackage{latexsym}

\usepackage{graphicx}
\usepackage{dcolumn}
\usepackage{amsmath}
\usepackage{amsfonts}
\usepackage{bm}

\newcommand{\be}{\begin{equation}}
\newcommand{\ee}{\end{equation}}
\newcommand{\bea}{\begin{eqnarray}}
\newcommand{\eea}{\end{eqnarray}}

\newcommand{\p}{\partial}
\newcommand{\s}{\sigma}

\newcommand{\la}{\langle}
\newcommand{\ra}{\rangle}
\newcommand{\rd}{\mbox{d}}
\newcommand{\ri}{\mbox{i}}
\newcommand{\re}{\mbox{e}}

\renewcommand{\vec}[1]{{\bm #1}}

\begin{document}
\title{A possible realization of multi-channel Kondo model in a system of magnetic chains}
\author{A. M. Tsvelik and Wei-Guo Yin}
\affiliation{ Department of Condensed Matter Physics and Materials Science, Brookhaven National Laboratory,
  Upton, NY 11973-5000, USA} \date{\today }
\begin{abstract}
We discuss a possible realization of overscreened multi-channel Kondo model in quasi-one-dimensional magnets. 

 \end{abstract}


\maketitle
\section{Introduction}

 Theory predicts \cite{frahm}, \cite{affleck} that isolated spins interacting with  spin S=1/2 Heisenberg chain can perform functions similar to magnetic impurities in metallic host. This points to  a new interesting physics; in particular it rises a prospect of realization  of the quantum critical behavior  existing in  the multichannel Kondo models \cite{tsvelik},\cite{andrei},\cite{ludwig}.

 The physical basis for the  similarity between  electron Kondo models and spin chain Kondo models comes from the fact that electronic charge plays no role in the Kondo phenomenon. The following arguments make this statement more rigorous. First, since local spin strongly scatters only one (or at best few) electron partical harmonics \cite{error}, the Kondo effect is essentially a one-dimensional phenomenon. As far as the impurity properties are concerned, the bulk electrons can be represented as  one-dimensional chiral ones  with a linear spectrum:
\bea
H_{bulk} = \sum_{|k|< \Lambda ,a} v_Fk\psi^+_a(k)\psi_a(k), \label{bulk}
\eea
where $\Lambda$ is the ultraviolet cut-off and $a$ index includes spin and channel degrees of freedom. Second,  one-dimensional Hamiltonian (\ref{bulk}) can be decomposed into the  commuting  charge and isotopic parts so that only the latter interact with the impurity spin (spin-charge separation). These two commuting Hamiltonians are expressed in terms of fermionic bilinears (currents) which compose closed algebras (Kac-Moody algebras). Third, the isotopic part of the bulk Hamiltonian is equivalent to the Hamiltonian describing a low energy excitations of a one-dimensional quantum magnet (see, for instance,\cite{wznw},\cite{wznw1}).

 A detailed study of the spin Kondo effect in a Heisenberg chain with an extra spin at the end is presented in \cite{affleck}. Earlier Affleck and Eggert described a possibility of two-channel Kondo effect \cite{eggert}. In \cite{affleck} the authors found a certain difference with the free electron Kondo effect due to the presence of marginally irrelevant bulk interactions. It turned out however,  that though these interactions  modify  the flow to the strong coupling fixed point, they do not change its character.

 Thus  we can be rest assured that the standard strong coupling fixed point can be realized in spin chains. We may wonder whether something more interesting can be achieved.  Since the pioneering prediction of a possibility of quantum critical point (QCP) in multi-channel Kondo model by Nozieres and Blandin \cite{blandin}, this QCP has never ceased to intrigue the researchers. However, its realization requires symmetry between the scattering channels which is very difficult to achieve in real materials, as was explained in the same paper \cite{blandin}. In the present paper we describe how these difficulties can be avoided in spin chains and outline particular examples of microscopic structures which may give rise to multichannel Kondo effect with number of channels greater than two. 

 The paper consists of two parts (besides the Introduction). In Section II we describe the model and provide its solution. In Section III we describe the quantum chemistry necessary for a realization of the model. 

\section{The model}

 Now we are going to apply the above  considerations to the model relevant to our case. This model consists of $N$ spin S=1/2 Heisenberg chains interacting with a cluster of $N$ extra spins 1/2:
\bea
&& H = \sum_{p=1}^N H_{0}^{(p)} + \sum_{p} {\cal J}_p{\bf S}_p[{\bf s}^p_{-1/2} + {\bf s}^p_{1/2}]\nonumber\\
&& -\sum_{p>q} J_{F,pq}{\bf S}_p{\bf S}_q\label{Model1}\\
&& H_0 = J\sum_{j=-L}^L{\bf s}_{j-1/2}{\bf s}_{j+1/2} \label{Heis}
\eea
The coupling constants are all positive $J, {\cal J}, J_F >0$ so that the interaction along the chains and the interaction between the  chain and the cluster spins are antiferromagnetic and the interactions within the cluster are ferromagnetic. It is important that the positive and negative sites of each chain interact symmetrically with the cluster (the 2nd term in (\ref{Model1})). In the next Section we will discuss how to achieve this arrangement in practice. It is also important that the exchange is isotropic. Besides these there is no other fine tuning in our model.

 The main statement of the paper is that for a certain range of parameters described below model (\ref{Model1}) displays the Quantum Critical behavior belonging to the universality class of the multichannel overscreened Kondo model.

 The starting point of our derivation is the well known non-Abelian bosonization of spin S=1/2 Heisenberg chain. The low energy limit of (\ref{Heis}) is given by the continuous limit Wess-Zumino-Novikov-Witten Hamiltonian \cite{affleck2}
\bea
&& H_{WZNW}^{(k)} = \label{wznw1}\\
&& \frac{2\pi v}{k+2}\int \rd x\Big(:J^a_R(x)J^a_R(x): + :J^a_L(x)J^a_L(x):\Big), \nonumber
\eea
with $k=1$ perturbed by a marginally irrelevant interaction
\bea
V_{pert} = - g'\int \rd x J^a_R(x)J_L(x), ~~ g' >0,  \label{pert}
\eea
$:A:$ means normal ordering of operator $A$ and $v = \pi J/2$ is the spinon velocity. The bare value of $g'$ can be reduced by introducing second neighbor exchange. Here $J_{R,L}$ are  the right and left current operators. The right and the left currents commute. Therefore in the limit of zero energies when the irrelevant interaction (\ref{pert}) dies out,  the excitations of the spin chain can be separated into left- and right- moving ones. The currents of the same chirality satisfy  the SU(2) Kac-Moody algebra of level $k$:
\bea
[J^a(x),J^b(y)] = \ri\epsilon^{abc}J^c(x)\delta(x-y) + \frac{\ri k}{4\pi}\delta_{ab}\delta'(x-y). \label{Kac}
\eea

 The continuum limit expression for spin operators of the Heisenberg chain is given by
\bea
&& {\bf s}_{j+1/2} = \label{spin}\\
&& {\bf J}_R(x) + {\bf J}_L(x) + \ri(-1)^j\mbox{Tr}\Big\{\vec\s[g^+(x)-g(x)]\Big\} +..., ~~ x = a_0j, \nonumber
\eea
where $a_0$ is the lattice constant and $g$ is an SU(2) matrix field of scaling dimension 1/2. This field is expressed nonlocally in terms of the current operators; its  more precise description is not necessary since due to the fine tuning of the interactions it does not contribute to the interaction with the cluster spins. The dots stand for less relevant operators. From (\ref{spin}) we see that
\bea
{\bf S}_p[{\bf s}^p_{-1/2} + {\bf s}^p_{1/2}] = 2{\bf S}_p\Big[{\bf J}_R(0) + {\bf J}_L(0)\Big]_p +...
 \eea
The continuum limit of (\ref{Model1}) is then
\bea
&& H = H_{bulk} + 2{\cal J}\sum_p {\bf S}_p\sum_{\alpha =R,L}{\bf J}_p^{\alpha}(0) - \sum_{p>q}J_{pq}{\bf S}_p{\bf S}_q \label{Model2}\\
&& H_{bulk} = \frac{2\pi}{3}\sum_p\sum_{\alpha = R,L}\int_{-\infty}^{\infty} \rd x :\Big({\bf J}^{\alpha}\Big)^2_p: + V_{pert} \label{wznw2}
\eea
The fundamental property of the current operators is that the sum of $k$ currents of level one produces a current operator of level $k$. The second fundamental property directly relevant to our calculation is the so called conformal embedding:
\bea
&& H_{WZNW}(k=1) + H_{WZNW}(k=1) =\nonumber\\
&&  H_{WZNW}(k=2) + \frac{v\ri}{2}\int \rd x \xi\p_x\xi, \label{embed},
\eea
where $\xi(x)$ is a real (Majorana) fermion field satisfying the anticommutation relations
\be
\{\xi(x),\xi(y)\} = 2\delta(x-y).
\ee
The third property is that $H_{WZNW}(k=2)$ and the currents can be written in terms of  three Majorana fermions
\bea
&& H_{WZNW}(k=2) = \frac{v\ri}{2}\sum_{a=1}^3\int \rd x \chi^a\p_x \chi_a, \nonumber\\
&&  J^a_{k=2} = \frac{\ri}{2}\epsilon^{abc}\chi^b\chi^c \label{Maj}
\eea
Since the right and the left currents commute and the impurity spins interact with them only at one point in space, the chirality can be treated as a channel index and if one neglects the marginally irrelevant bulk interaction, than  model (\ref{Model2}) can be rewritten in terms of the chiral fermions:
\bea
&& H = \sum_p H_{2-channel}^{(p)} - \nonumber\\
&& \sum_{p>q}J_{pq}{\bf S}_p{\bf S}_q + \sum_p\frac{v\ri}{2}\int \rd x \xi_p\p_x\xi_p,
\eea
where $\xi$ does not take part in the interaction with the impurity spins. The 2-channel Kondo Hamiltonian is \cite{ioffe}:
\bea
H_{2-channel} = \frac{v\ri}{2}\sum_{a=1}^3\int \rd x \chi^a\p_x \chi^a + \ri {\cal J}S^a\epsilon^{abc}\chi^b(0)\chi^c(0). \label{2chan}
\eea

 Now we will consider different limits. One possibility is to have a strong ferromagnetic interaction so that total spin of the cluster is $S=N/2$. Then the bulk effectively interacts only with this spin with a reduced exchange coupling \cite{schrieffer}:
\bea
H_{eff} = \frac{1}{N}{\bf S}\sum_{p}{\cal J}_p{\bf J}_{k=2}^{(p)} + \frac{v\ri}{2}\sum_p\sum_{a=1}^3\int \rd x \chi^a_p\p_x \chi^a_p
\eea
The multichannel fixed point appears only if ${\cal J}_p$'s are equal. Then since the sum of $N$ $k=2$ currents is equivalent to one $k=2N$ current which is enough to overscreen spin $S=N/2$, we will  get a non-Fermi liquid ground state of the $2N$-channel Kondo model with the Kondo temperature
\bea
T_K^{strong} \sim J\exp[- N\pi v/{\cal J}]. \label{TKstrong}
\eea
The condition for the validity of this approach is $J_F >> T_K^{strong}$.

 The other limit is the one of a weak intra-cluster exchange. In that case the Kondo screening proceeds in two stages. First we consider the screening of individual spins. This occurs in the framework of the 2-channel Kondo model (\ref{2chan}). This model is quantum critical and exactly solvable \cite{tsvelik},\cite{andrei},\cite{ludwig}. For energies below the Kondo temperature
\bea
T_K\sim {\cal J}\exp(- v\pi/2{\cal J})
\eea
 the effective action renormalizes into
\bea
&& S_{eff} = \frac{1}{2}\sum_{a=1}^3\int \rd x \chi^a(\p_{\tau} - \ri v\p_x )\chi_a + \frac{\ri}{2}\eta\p_{\tau}\eta +\nonumber\\
&&  \tilde g\eta\chi_1(0)\chi_2(0)\chi_3(0), \label{qcp}
 \eea
$\epsilon$ is a local zero energy Majorana mode residing on the impurity site $x=0$. The last term represents the leading irrelevant interaction. As was shown in \cite{ioffe}, at the 2-channel Kondo model QCP the local spin $S^a$ renormalizes into
\bea
\hat S^a \rightarrow \ri {T_K}^{-1/2}\eta \chi^a(0) \label{spin3}
\eea
 As was shown in \cite{affleck}, the presence of the marginally irrelevant bulk interaction may modify the expression for $T_K$, but the critical point itself remains intact.

 As we have said, we consider the case  $T_K >> J_{pq}$. Then the 2-channel QCP is reached first and the spin-spin interaction in the cluster can be considered as a perturbation. Using (\ref{spin3}) we obtain
\bea
-J_{pq}{\bf S}_p{\bf S}_q \rightarrow -\frac{J_{pq}}{T_K}\eta_p\eta_q\sum_{a=1}^3\chi_p^a(0)\chi_q^a(0). \label{Int}
\eea
Now unlike the strong exchange limit it seems there is a difference between different cluster sizes. For $N=2$ the interaction (\ref{Int}) becomes
\bea
V = - J_{12}\eta_1\eta_2 \sum_{a=1}^3\chi_1^a(0)\chi_2^a(0) \label{Pert1}
\eea
Since $\ri\eta_1\eta_2$ is just a Pauli matrix $\s^z$ and it commutes with the Hamiltonian (provided we neglect the irrelevant operators in (\ref{qcp})), we can replace it by $+ 1$ or $-1$. The resulting perturbation (\ref{Pert1}) is exactly marginal and does not flow. This is different from the strong exchange case where for $N=2$ we got 4-channel Kondo effect.

Now consider $N=3$. The Majorana zero mode bilinears naturally combine into components of spin 1/2 operator \cite{martin}:
 \be
T^a = \frac{\ri}{2}\epsilon^{abc}\eta^b\eta^c, ~~ \{\eta^a,\eta^b\} = 2\delta^{ab},
\ee
and the interaction (\ref{Int}) becomes
\bea
 V = \gamma_{p}T^p\sum_a J^p_a(0),
\eea
where $\gamma_p >0$ are related to $J_{pq}$ and
 and $J_p^a = \frac{\ri}{2}\epsilon^{abc}\chi_b\chi_c$ are level $k=2$ Kac-Moody currents. Since the sum of three $k=2$ currents is $k=6$ Kac-Moody current ${\cal F}^a$ the effective action for our cluster model (\ref{Model1})  at energies $<< T_K$ is
\bea
&& H = H_{bulk} + \gamma_{p}T^p{\cal F}^p(0), \label{6chan}\\
&& H_{bulk} = H_{WZNW}[SU(2),k=6] +\nonumber\\
&&  H_{WZNW}[SU(6),k=2] \nonumber,
\eea
where
\bea
H_{WZNW}[SU(2),k=6] = \frac{\pi v}{4}\int \rd x :{\cal F}^a{\cal F}^a:.
\eea
Thus for $N=3$ we get the same result both for strong and weak intra-cluster exchange.

The exact solution for $N>3$ is absent, but seems likely that we will still get QCP.

It is remarkable that the QCP is robust with respect to a possible difference between intra-cluster exchange constants $J_{pq}$. However, it is quite sensitive to the channel anisotropy. The latter one may come from two sources. First, it appears if the SU(2) symmetry of the exchange interactions is broken which may occur for strong spin orbit interactions. The second possible source is an asymmetry in the interaction of the impurity spins.

\section{Quantum Chemistry}

\begin{figure}[h]
\includegraphics[width=1.0\columnwidth]{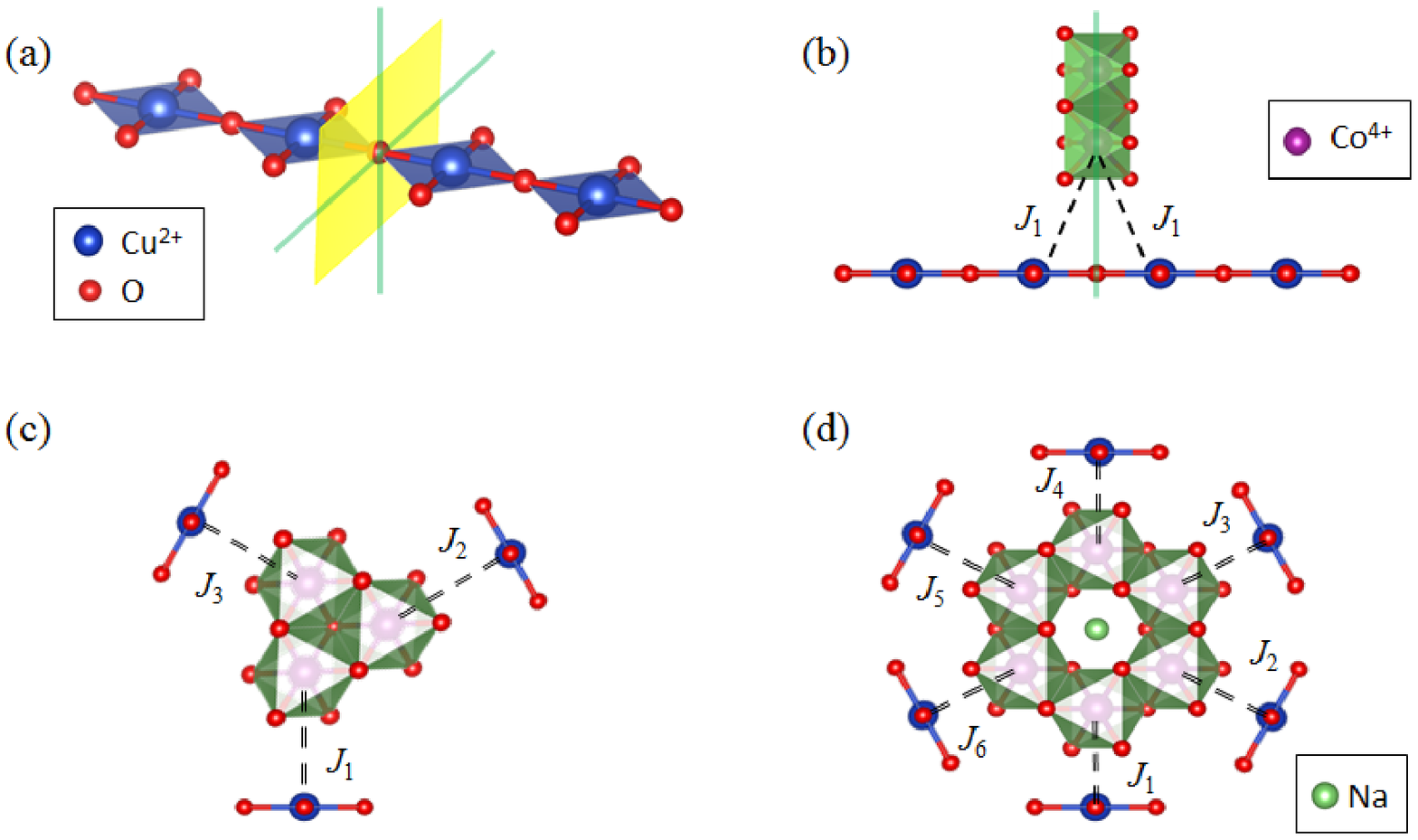}
\caption{An illustration of possible structural configurations of the model (2). (a) A CuO$_2$ spin-1/2 Heisenberg chain with the yellow plane (color online) being a mirror symmetry plane with  two high symmetry lines (green color online). (b) The side view of a spin-1/2 Heisenberg chain coupled by a triangular CoO$_2$ cluster allocated to  one of the high symmetry lines. (c) Three chains coupled by a triangular CoO$_2$ cluster viewed along the chain direction. (d) A 6-chain system coupled by a 6-spin cluster. ${\cal J}_p$ is the antiferromagnetic exchange coupling between the $p$th chain and the cluster. The ferromagnetic interaction within the cluster might be realized by the $90^\circ$ Co-O-Co exchange pathways \cite{Goodenough,Kanamori}.}
\label{fig:illus}
\end{figure}

Now we briefly address some practical issues concerning a possible materialization of the model (2). The task is  challenging, since the aforementioned QCP requires symmetry between the scattering channels  and thus special symmetries of the two components: the chains and the cluster. Therefore, at most we can do is to present a case  that this can be done in principle. As an example of the  spin-1/2 quantum Heisenberg chain we take  CuO$_2$ chain as in Sr$_2$CuO$_3$ [see Fig.~\ref{fig:illus}(a)]. For the $N=1$ two channel Kondo model, the requirement that both halfs  of each chain interact symmetrically with a single-site magnetic impurity (the ${\cal J}_{p}$ term in (\ref{Model1})) may be achieved by allocating the impurity to a mirror plane of the chain (the yellow plane in Fig.~\ref{fig:illus}(a)). However, for a multi-site cluster, symmetry consideration is significantly strengthened---it is very likely that this condition can be satisfied only when the ``contact'' magnetic ion of the cluster is allocated to the two high symmetry lines in the mirror plane (the green lines in Fig.~\ref{fig:illus}(a)), as shown in Fig.~\ref{fig:illus}(b). The good news is that these high symmetry lines usually pass trough the center of the interstitial areas.

As for the ferromagnetic interaction within the cluster, one may resort to the Goodenough-Kanamori rule \cite{Goodenough,Kanamori} to use the $90^\circ$ $M$-O-$M$ exchange pathways, where $M$ stands for the spin-1/2 magnetic ion in the cluster. To reach QCP in the large $J_{pg}$ limit, the number of $M$, $n_M$, needs to satisfy $n_M<2N$ \cite{blandin}. Therefore, the cluster has to be compact in space. For $N=3$, a possible realization is a triangular CoO$_2$ cluster (see Fig.~\ref{fig:illus}(c)), where the cobalt ions sit inside the edge-sharing oxygen octahedra, forming the $90^\circ$ Co-O-Co exchange pathways, and the $d$ electrons on the Co$^{4+}$ sites are in the spin-1/2 $t^{5}_{2g}$ configuration. The three $t_{2g}$ orbitals ($xy$, $yz$, $zx$) are related mutually by rotation around the cubic $x$, $y$, or $z$ axis and thus can sustain the triangular structure \cite{Jackeli}. The oxygen-mediated ferromagnetic interaction is usually accompanied by the direct $M$-$M$ antiferromagnetic superexchange coupling, which leads to the antiferromagnetic states in Na$_x$CoO$_2$ crystals. However, in a cluster the Hubbard interaction ($U$) is expected to be significantly larger and thus the direct superexchange (which is roughly proportional to $1/U$) would be unfavored. Fig.~\ref{fig:illus}(d) shows a larger cluster that may accommodate $N=4$ to $6$.

In addition, as shown in Figs.~\ref{fig:illus}(c) and ~\ref{fig:illus}(d), there is a large space between the chains that needs to be stuffed. It seems that the symmetry set up by the chains and the cluster can be preserved by a symmetric stuffing on the left and right of the mirror plane of the chains. The details of how to stuff is beyond our reach.

\section{Conclusions and Acknowledgements}

 We have demonstrated a feasability of realization of the multichannel Kondo effect in systems of antiferromagnetic chains coupled to solitary spins or clusters of spins.  

 The work at Brookhaven National Laboratory was supported by the U.S. Department of Energy (DOE), Division of Materials Science, under Contract No. DE-AC02-98CH10886.


\begin{thebibliography}{99}
\bibitem{error} The literature is peppered with persistent statements that this harmonics is the $s$-wave one. This is wrong. The impurity spin made of $d$ ($f$) -electrons interacts with the $d$ ($f$) partial harmonics of conduction electrons.
\end{thebibliography}
\end{document}